\documentclass[twocolumn,aps,prl]{revtex4}

\usepackage{graphicx,amsmath,amssymb}

\begin{document}
\title{Reply to Comment on ``Replica symmetry breaking in trajectories of a driven Brownian particle''}
\author{Masahiko Ueda}
\email[]{ueda.masahiko.5r@kyoto-u.ac.jp}
\address{Department of Systems Science, Graduate school of Informatics, Kyoto University, Kyoto 606-8501, Japan}
\author{Shin-ichi Sasa}
\email[]{sasa@scphys.kyoto-u.ac.jp}
\address{Department of Physics, Graduate school of Science, Kyoto University, Kyoto 606-8502, Japan}


\maketitle

In their Comment \cite{SinBar}, Singha and Barma study the probability distribution $P_t(d)$ of relative distance $d$ between two particles on a common Kardar-Parisi-Zhang (KPZ) potential field at time $t$.  They present
numerical evidence for the claim that a scaling form of $P_t(d)$ exhibits
a power-law behavior in the short-distance regime, not a stretched
exponential behavior as numerically found in Ref. \cite{UedSas2015}.  Their scaling
form is also in accordance with our results that (i) the mean-squared
relative distance is normal $\left\langle d^2 \right\rangle \sim t$, and
(ii) the peak $P_t(d=0)$ is time-independent, reported in our
Letter \cite{UedSas2015}. Furthermore, using the scaling from, they explain
our result that (iii) the probability distribution $P_t(d)$ decays to zero
for the Edwards-Wilkinson (EW) potential fields.
Their results are reasonable, and we do not have any objection to their
comment. 

\begin{figure}[bp]
\includegraphics[clip, width=8.0cm]{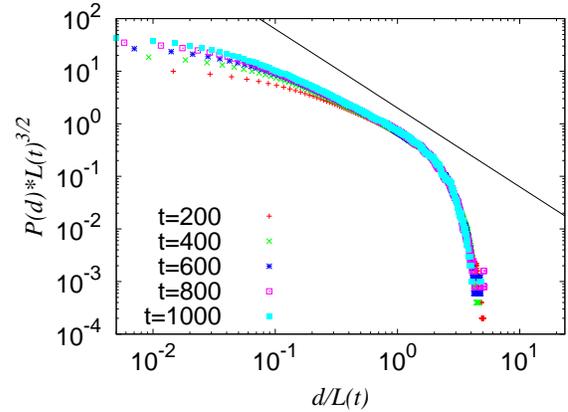}
\caption{The probability distribution $P_t(d)$ in the scaling form for the KPZ case. The black line represents the slope $-3/2$.}
\label{fig:Pd_KPZ}
\end{figure}

\begin{figure}[bp]
\includegraphics[clip, width=8.0cm]{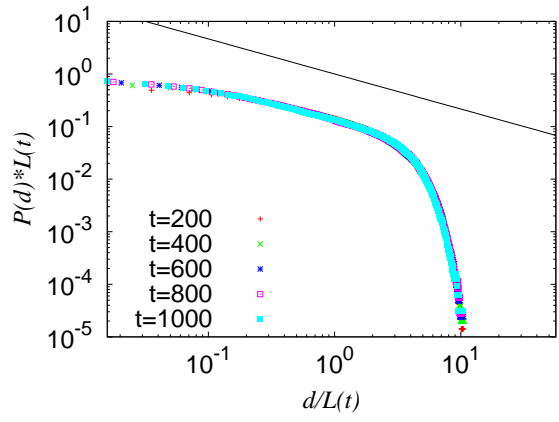}
\caption{The probability distribution $P_t(d)$ in the scaling form for the EW case. The black line represents the slope $-2/3$.}
\label{fig:Pd_EW}
\end{figure}

We here point out that the model studied in Ref. \cite{SinBar} is not
the exactly same as that in our Letter \cite{UedSas2015}.
In Ref. \cite{UedSas2015}, particles are driven by  the KPZ surface
in continuous space with cutoff length $\Delta x$.
In our model, the Langevin equation for particles and the KPZ equation for surface are numerically solved.
In contrast, in Ref. \cite{SinBar}, particles are defined on a discrete lattice.
In their model, hills and valleys of the surface are stochastically changed to each other, and particles move downward or upward stochastically \cite{SinBar2018}.
Although long-time and long-distance behavior
of the two models is expected to be the same, microscopic
details of the models may make a crucial difference in short-distance
behavior. In fact, we plot our numerical data by using the scaling form
proposed in Ref. \cite{SinBar} in Fig. \ref{fig:Pd_KPZ}.
The data for $d/\mathcal{L}(t)\simeq 1.0$ with coarsening length scale $\mathcal{L}(t)\sim t^{2/3}$ collapse and are reasonably fitted by a power-law function with slope $-3/2$, while  the data for smaller $d/\mathcal{L}(t)$ do not collapse.
This may be because the cutoff $\Delta x = 0.5$ makes an effect on short-distance behavior. 
Since we use linear interpolation for calculation of potential fields, and meaningful $d$ is $d\gg \Delta x$, data for $d/\mathcal{L}(t) \geq 10\Delta x/200^{2/3} \simeq 0.15$ should be investigated for $200\leq t \leq 1000$.
This fact can also be seen from the EW case in Fig. \ref{fig:Pd_EW} with $\mathcal{L}(t)\sim t^{1/2}$.
Whereas data collapse is better for the EW case, the slope deviates from their prediction for $d/\mathcal{L}(t)\leq 10^{-1}$.

In sum, their numerical finding of the scaling form for their model supports that replica symmetry breaking (RSB) in trajectory space is observed even if the short-distance behavior is different.
Since RSB in trajectory space is related to the fact that $\lim_{t\rightarrow \infty}P_t(d=0)>0$, their result confirms RSB for the KPZ case by the existence of a scaling function.


\begin{thebibliography}{99}
 \bibitem{SinBar} T. Singha and M. Barma, ``Comment on ``Replica symmetry breaking in trajectories of a driven Brownian particle''''.
 \bibitem{UedSas2015} M. Ueda and S.-i. Sasa, Phys. Rev. Lett. {\bf 115}, 080605 (2015).
 \bibitem{SinBar2018} T. Singha and M. Barma, Phys. Rev. E {\bf 97}, 010105(R).
\end{thebibliography}
\end{document}